\documentclass[11pt]{article}
\usepackage{moriond,epsfig}

\def\be{\begin{equation}}
\def\ee{\end{equation}}

\newcommand{\ba}{\begin{eqnarray}}
\newcommand{\ea}{\end{eqnarray}}
\newcommand{\dis}{\displaystyle}
\newcommand{\tr}{\mbox{tr}}
\newcommand{\re}{\mbox{Re }}
\newcommand{\im}{\mbox{Im }}

\newcommand{\barr}{\begin{array}{c}}
\newcommand{\earr}{\end{array}}

\begin{document}
\begin{flushright}
BUTP-2003-11\\
CAFPE-18-03\\
UGFT-148-03\\
hep-ph/0305164
\end{flushright}
\vspace*{4cm}
\title{CP--VIOLATING ASYMMETRIES IN $K^+\to 3\pi$ IN THE STANDARD MODEL}

\author{ELVIRA  G\'AMIZ and JOAQUIM PRADES}
\address{CAFPE and Departamento de F\'{\i}sica Te\'orica y del Cosmos\\
Universidad de Granada, Campus de Fuente Nueva, E--18002 Granada, Spain}
\author{IGNAZIO SCIMEMI}
\address{Institute of Theoretical Physics, University of Bern, 
Sidlerstr. 5, CH--3012 Bern, Switzerland}

\maketitle
\abstract{
\noindent We  update the CP--violating  asymmetries in $K^+\to 3 \pi$. 
In particular we study $\Delta g_C$ and $\Delta g_N$
--the  asymmetries in the slope $g$. We emphasize its complementarity
to the direct CP--violation parameter $\varepsilon_K'$ and
 the large sensitivity to the size of the 
imaginary part of the octet coupling $\im G_8$ at lowest order in CHPT.
We also give the prospects of  a full calculation
 at next--to--leading order 
of  this and other CP--violating asymmetries
which will be presented elsewhere.}

\vspace*{0.5cm}
We have calculated \cite{GPS03} 
\ba
\label{defdecays}
K_2(k)&\to&\pi^0(p_1)\pi^0(p_2)\pi^0(p_3)\,,\nonumber\\
K_2(k)&\to&\pi^+(p_1)\pi^-(p_2)\pi^0(p_3)\,,\nonumber\\
K_1(k)&\to&\pi^+(p_1)\pi^-(p_2)\pi^0(p_3)\,,\nonumber\\
K^+(k)&\to&\pi^0(p_1)\pi^0(p_2)\pi^+(p_3)\,,\nonumber\\
K^+(k)&\to&\pi^+(p_1)\pi^+(p_2)\pi^-(p_3)\,,
\ea
as well as their  CP--conjugated decays 
at one--loop in Chiral Perturbation Theory (CHPT) \cite{WEI79,GL84,GL85} 
--see \cite{schools} for reviews.
 Above, the momentum of each particle is denoted between brackets  and 
\be 
K_{1(2)} = \frac{K^0-(+)\overline{K^0}}{\sqrt 2} \, .
\ee
 The calculation has been done in the limit of equal up and down light 
quark masses and we have 
included one--loop electromagnetic corrections for the 
dominant  octet part. We have not included photon loops.
 
CHPT is the effective field theory of the Standard Model at energies 
below one GeV . The lowest order SU(3) $\times$ SU(3)
 chiral Lagrangian describing $|\Delta S|=1$ transitions is
\ba
\label{deltaS1}
{\cal L}^{(2)}_{|\Delta S|=1}&=&
C\,F_0^6 \, e^2 \, G_E \, \tr \left( \Delta_{32} u^\dagger Q u\right)
+ C F_0^4 \left[ G_8 \, \tr \left( \Delta_{32} u_\mu u^\mu \right)
+ G_8' \tr \left( \Delta_{32} \chi_+ \right) \right.
\nonumber \\ 
&+& \left. 
G_{27} \, t^{ij,kl} \, \tr \left( \Delta_{ij} u_\mu \right) \,
\tr \left(\Delta_{kl} u^\mu\right) \right] + {\rm h.c.}\ ,
\ea
with
\be
C= -\frac{3}{5} \frac{G_F}{\sqrt 2} V_{ud} {V_{us}}^* \simeq -1.07 \cdot 
10^{-6} \, {\rm GeV}^{-2} .
\ee
$F_0= (87 \pm 6 )$ MeV  is the chiral limit value of the pion decay
constant $f_\pi= (92.4 \pm 0.4)$ MeV, 
\ba
u_\mu \equiv i u^\dagger (D_\mu U) u^\dagger = u_\mu^\dagger \; , 
\nonumber \\
\Delta_{ij}= u \lambda_{ij} u^\dagger\; , \nonumber \\
(\lambda_{ij})_{ab}\equiv
\delta_{ia} \delta_{jb}\; , \nonumber \\ 
\chi_{+(-)}= u^\dagger \chi u^\dagger +(-)u \chi^\dagger u  \ ,
\ea 
$\chi= \mbox{diag}(m_u,m_d,m_s)$ ,  
a 3 $\times$ 3 matrix collecting the light quark masses,
and  
$U\equiv u^2=\exp{(i\sqrt 2 \Phi /F_0)}$ is the exponential
representation incorporating the octet of light pseudo-scalar mesons
in the SU(3) matrix $\Phi$, 

\ba
\Phi\equiv \left(  
\begin{array}{ccc}
\frac{\dis \pi^0}{\dis \sqrt{2}} + 
\frac{\dis \eta_8}{\dis \sqrt{6}} & \pi^+ & K^+ 
\nonumber \\ 
\pi^- & -\frac{\dis \pi^0}{\dis \sqrt{2}} 
+ \frac{\dis \eta_8}{\dis \sqrt{6}} & K^0 
\nonumber \\ 
K^- & \bar{K^0} &- 2 \frac{\dis \eta_8}{\dis \sqrt{6}}  
\end{array}
\right) .
\ea

\noindent
The non--zero components of the 
SU(3) $\times$ SU(3) tensor $t^{ij,kl}$  are
\ba
t^{21,13}=t^{13,21}=\frac{1}{3} , \, &  t^{22,23}=t^{23,22}=
-\frac{\dis 1}{\dis 6} \, ,
\nonumber \\
t^{23,33}=t^{33,23}=-\frac{1}{6} , \, 
&  t^{23,11}=t^{11,23}=\frac{\dis 1}{\dis 3},
\ea
and $Q=\mbox{diag}(2/3,-1/3,-1/3)$ is a 3 $\times$ 3
matrix which collects the electric charge of the three light 
quark flavors. The Lagrangian needed to describe $K\to 3\pi$
at next--to--leading order 
can be found in \cite{GL84,GL85,KMW90,EF91,EKW93,EIMNP00}.

 There is a long history of work concerning $K\to 3 \pi$. First, using
current algebra and  tree--level chiral Lagrangians \cite{HOL69}.
The one--loop calculation was  done in \cite{KMW91}, 
unfortunately the analytical full results were not available.
Recently there has appeared the first  full published result in 
\cite{BDP03}. We have confirmed this result 
redoing the one--loop calculation \cite{GPS03}. 

Using the $K \to 3 \pi$  one--loop results and the ones in \cite{BPP98} 
for $K\to \pi \pi$,
a fit to all measured CP--conserving amplitudes and decay rates  of those 
two transitions was done in \cite{BDP03} updating the results 
 in \cite{KMW91}. 
The result found  there for the ratio of the isospin
definite [0 and 2 ] $K\to\pi\pi$ amplitudes  to all orders in CHPT
is 
\be
\frac{A_0[K\to\pi\pi]}{A_2[K\to\pi\pi]}= 21.8\ ,
\ee
giving the infamous $\Delta I=1/2$ rule  and
\be
\left[\frac{A_0[K\to\pi\pi]}{A_2[K\to\pi\pi]}\right]^{(2)}= 17.8 \ ,
\ee
to lowest CHPT order $p^2$. I.e., Final State Interactions (FSI)
and the rest of higher order corrections account for 22 \% of the
$\Delta I=1/2$ rule \cite{BDP03}.  Yet most of this enhancement
appears at lowest CHPT order !
The last result is equivalent to the values of the lowest order 
coupling constants
\ba
\label{exp}
\re G_8 = 5.5 \pm 0.5  & {\rm and} & G_{27}= 0.39 \pm 0.05. 
\ea
In this normalization, $G_8=G_{27}=1$ at large $N_c$.

CP--violating observables in  $K\to 3 \pi$ has attracted also
a lot of work since long time ago 
\cite{LW80,AVI81,GRW86,DHV87,BBEL89,DIP91,SHA93,DIPP94}
 and references therein.

In this talk, we  update on the CP--violating asymmetries in the 
slope $g$ defined as
\ba
\frac{\left|A_{K\to3\pi}(s_1,s_2,s_3)\right|^2}
{\left|A_{K\to3\pi}(s_0,s_0,s_0)\right|^2}= 1 + g \, y + { \cal O}
(x^2, y^2)\  ,
\ea
with $s_i\equiv (k-p_i)^2$, $3 s_0 \equiv m_K^2 + m^2_{\pi^{(1)}}+
m^2_{\pi^{(2)}}+m^2_{\pi^{(3)}}$ and the Dalitz variables
\ba
x\equiv \frac{s_1-s_2}{m^2_{\pi^+}} , & 
y \equiv \frac{\dis s_3-s_0}{\dis m^2_{\pi^+}} .
\ea

The two asymmetries that we will be discussing here are
\ba
\Delta g_C \equiv \frac{g[K^+\to \pi^+ \pi^+ \pi^-]-
g[K^-\to\pi^-\pi^- \pi^+]}
{g[K^+\to \pi^+ \pi^+ \pi^-]+g[K^-\to\pi^-\pi^- \pi^+]}
\ea
and
\ba
\Delta g_N \equiv \frac{g[K^+\to \pi^0 \pi^0 \pi^+]-
g[K^-\to\pi^0\pi^0\pi^-]}
{g[K^+\to \pi^0 \pi^0 \pi^+]+g[K^-\to\pi^0\pi^0 \pi^-]} .
\ea

They have been discussed in the literature before finding
conflicting results 
\be
\left|\Delta g_C\right|\simeq \left| \Delta g_N \right| 
\simeq 11.0 \times 10^{-4}\ ,
\ee
in \cite{BBEL89} ,
\be
\left| \Delta g_C \right| \simeq 0.07 \times 10^{-4} \, , 
\ee
in \cite{DIP91},  and
\be
\left| \Delta g_C \right| \simeq 0.016 \times 10^{-4} \ ,
\ee 
in \cite{IMP92}. The first result was claimed to be at one--loop, however
they did not used CHPT fully at one--loop. 
The last two results are at lowest order $p^2$, but they made assumptions
to get those results too. 
Among them,  they assumed  the exact dominance of the
gluonic penguin and  neglected the NLO chiral corrections
in the result of $\varepsilon'_K$ to extract 
the contribution of the gluonic penguin to
the asymmetries.   They also made some estimate 
of the NLO based on strong assumptions   which could increase
their LO result up to one order of magnitude.

 Recently NA48 has announced the possibility of measuring
the asymmetry $\Delta g_C$  with a sensitivity of the order of $10^{-4}$ 
and maybe $\Delta g_N$, see for instance \cite{NA48}.
It is therefore mandatory to have an update of 
these predictions  at  NLO in CHPT. 

We present here the CHPT leading order results  for those asymmetries.
The complete NLO analysis including the dominant contributions to the
imaginary part at the NLO non--trivial order, i.e. at order $p^6$,
  will be ready soon \cite{GPS03}.
At LO, the CP--violating asymmetry in the slopes $\Delta g_{C(N)}$ 
can be written  to a very good approximation as
\ba
\Delta g_{C(N)}  \simeq \frac{m_K^2}{F_0^2} 
\, A_{C(N)}  \, \im G_8 +  B_{C(N)} \, \im (e^2 G_E)\ ,
\ea
with 
\ba
A_{C(N)}=f_{C(N)}(\re G_8, G_{27}, m_K^2, m_\pi^2) &  {\rm and}  &
B_{C(N)}=h_{C(N)}(\re G_8, G_{27}, m_K^2, m_\pi^2) .
\ea
The analytic formulas can be found in \cite{GPS03}.

Numerically, using the results in (\ref{exp})  for $\re G_8$ and $G_{27}$, 
\ba
\Delta g_C \simeq 0.015 \, \im G_8 - 0.0014 \, \im (e^2 G_E) \, ,
\nonumber \\
\Delta g_N \simeq -0.0065 \, \im G_8 - 0.0008 \, \im (e^2 G_E) \, .
\ea


Let us see what  we do know about  the couplings $\im G_8$ and $\im G_E$.
At large $N_c$,  the contributions to $\im G_8$ and $\im G_E$ are
factorizable and  there is no scheme nor matching  scale dependence .
The unfactorizable topologies bring   unrelated dynamics, so that 
we cannot  give any uncertainty to the large $N_c$ result.
 We get
\ba
\im G_8\Big|_{N_c} = (2.0 \pm ? ) \,  \im \tau \, , \nonumber \\
\im (e^2 G_E)\Big|_{N_c} =  - (4.3 \pm ? ) \, \im \tau \, . 
\ea
In the Standard Model
\be
\im \tau \equiv - \left| \frac{V_{td}V_{ts}^*}{V_{ud}V_{us}^*} \right| 
\simeq -6.05 \times 10^{-4} ,
\ee
 and we used  here  \cite{BPR95}
\be
\langle \overline q q \rangle_{\overline{\rm{MS}}}
(2 {\rm GeV}) = - (0.018\pm0.004) {\rm GeV}^3 ,
\ee
which agrees with most recent results for the quark condensate
and light quark masses from QCD sum rules and light quark masses from 
 lattice QCD, see \cite{GPS03} for references.
At this order, 
\ba
\Delta g_C\Big|^{LO}_{N_c}  \simeq - 0.22 \times 10^{-4} \, & {\rm and} &
\Delta g_N\Big|^{LO}_{N_c}  \simeq  0.57 \times 10^{-5} \, .
\ea
There have been recently advances on going beyond the leading
order in $1/N_c$ in both couplings, $\im G_8$ and $\im G_{E}$.

In \cite{CDGM03,NAR01,BGP01}, 
there are recent model independent calculations of $\im G_E$.
 The results  there  are valid to all orders in $1/N_c$ and NLO
in $\alpha_S$. They are obtained using  hadronic tau data collected 
by the ALEPH and OPAL Collaborations at LEP. 
The agreement is quite good between them
and their result can be summarized in 
\be
\label{Q8}
\im (e^2 G_E)   = - (5.9 \pm 1.3 )  \, \im \tau\ , 
\ee
where the central value is an average and the error is the smallest one.
In \cite{KPR01} it was used a Minimal Hadronic Approximation 
to large $N_c$ to calculate $\im G_E$, they get
\be
\im (e^2 G_E) = -(9.9 \pm 3.0 ) \, \im \tau ,
\ee
which is also in agreement though somewhat larger.
There are also lattice results for $\im G_E$ 
using domain--wall fermions \cite{domwall}
 and  Wilson fermions \cite{wilson}.
All of them made the chiral limit extrapolations, their results
are in agreement between them and the average gives
\be
\im (e^2 G_E)  = -(4.7 \pm 0.4) \, \im \tau \, .  
\ee

There are also results  about $\im G_8$ at NLO in $1/N_c$.
In \cite{BP00}, the authors did a calculation using a hadronic
model which reproduced the $\Delta I = 1/2$ rule through
very large $Q_2$ penguin--like contributions \cite{BP99}
\ba
\re G_8 = 5.9 \pm 1.5 , & {\rm and} &
G_{27} = 0.33 \pm 0.10 \, , 
\ea  
 in very good agreement with the experimental results in (\ref{exp}).
The result  found there was at NLO in $1/N_c$ 
\be
\label{Q6}
\im G_8  = (5.1 \pm 1.2) \, \im \tau \, . 
\ee
The hadronic model used there had however some drawbacks
\cite{PPR98} which will be eliminated and the work eventually updated
following the lines in \cite{BGLP03}.

Very recently, using a Minimal Hadronic Approximation to
large $N_c$, the authors of \cite{HPR03} found qualitatively similar
results. I.e. enhancement toward the explanation of the $\Delta I=1/2$
rule  through $Q_2$ penguin--like diagrams and a matrix element
of the gluonic penguin $Q_6$ around  three times the factorizable
contribution. 

After these results  we can safely neglect the EM penguin contribution
to $\Delta g_{C(N)}$ and  conclude 
that these asymmetries are to a very good
approximation proportional to $\im G_8$. This is interesting since
with a good enough accuracy, this asymmetry can check a large
$\im G_8$.
This makes also this asymmetry complementary to the direct CP--violating
parameter $\varepsilon_K'$ where there is a strong cancellation
between the $\im G_8$ and $\im G_E$ contributions.

The results we get at LO are
\ba
\Delta g_C\Big|_{{\rm{LO}}} = -\left[ 0.51 \pm 0.25\right] \times 10^{-4}
 & {\rm and} &
\Delta g_N\Big|_{{\rm{LO}}} = \left[ 1.70 \pm 0.85\right] \times 10^{-5} \ ,
\ea
where we have used the results  in (\ref{Q8}) and (\ref{Q6}).

At NLO,  we have to include several ingredients.
For the real part, we need the chiral logs and counterterms, $K_i$.
The chiral logs  are analytically known \cite{BDP03,GPS03}.
The real part of the 
counterterms, $\re K_i$, that enter in $K\to 3 \pi$ can be obtained
from the fit to data done in \cite{BDP03}.

At NLO we also need $\im G_8'$ in addition to $\im G_8$,
for which,  to the best of our knowledge, 
there is just one calculation at NLO in $1/N_c$ at 
present in \cite{BP00}.  The results found there,  using
the same hadronic model discussed above, are
\ba
\re G_8' = 0.9 \pm 0.1 , & {\rm and} &
\im G_8' = (1.0\pm0.2) \im \tau \, .
\ea
The imaginary part of the rest of the counterterms, $\im K_i$, 
is much more problematic
and there is no NLO in $1/N_c$ calculation of them at present.

We can use several approaches to get the order of magnitude
and/or the signs of the rest of the imaginary parts
of the purely order $p^4$  couplings. 
 These approaches to get  information of the
those  unknown imaginary parts are  factorization plus meson dominance
\cite{EGPR89}, 
large $N_c$, or assuming that the ratio of the real   and imaginary parts
are dominated by the same dynamics at all orders in CHPT, therefore  
\be
\frac{\im K_i}{\re K_i} 
\simeq \frac{\im G_8}{\re G_8} \simeq \frac{\im G_8'}{\re G_8'}.
\ee
Notice that the last relation is well satisfied by the model 
calculation in \cite{BP00}.
This will allow us to check 
the counterterm dependence of the CP--violating asymmetries.
 
Using the estimates of the imaginary part of the counterterms above
and the results from the fit in \cite{BDP03} for the real part,
we have checked that the 
sum $g(K^+\to \pi^+\pi^+\pi^-)+g(K^-\to\pi^-\pi^-\pi^+)$
is dominated by chiral logs.  In the difference 
$g(K^+\to \pi^+\pi^+\pi^-)-g(K^-\to\pi^-\pi^-\pi^+)$ the counterterm
 contribution is however important  and needs more work.

The last ingredient to have a complete analysis of the CP--violating
asymmetries at NLO are the phases generated by the two--loop 
amplitudes $K\to 3 \pi$. The lowest
non-trivial order is $p^4$ which is analytically known. This is 
the one that has to be 
used in LO prediction of $\Delta g_C$. At NLO, one needs the 
same calculation at
order $p^6$ which is not available at present. However, one can 
do a very good
job using the known results at order $p^4$ and the optical theorem
to get analytically the  order $p^6$ imaginary parts that come
from two one--loop bubbles which we can assume to be dominant 
\cite{GPS03,DIPP94} to a very good accuracy. 
 We disregard three-body rescattering  since they  
cannot be written as a bubble resummation. 
One can expect them to be rather small
being suppressed by  the phase space available.
 
We have presented here an update of the $K^+ \to 3 \pi$ 
CP--violating asymmetries $\Delta g_{C(N)}$ at LO in the Standard Model.
We are working  in the full NLO corrections which 
will be ready soon \cite{GPS03}.

The measurement of this decay by NA48 or elsewhere will 
be extremely interesting
for several reasons.
We have seen that the asymmetries $\Delta g_{C(N)}$ 
are proportional at lowest order to $\im G_8$ to a very good approximation.
This allows a consistency check between the theoretical predictions
for $\varepsilon_K'$ \cite{BP00,HPR03,PPI01}.  Any prediction
for $\varepsilon_K'$ has to be also able to predict the CP--violating
asymmetries.  The measurement of those
asymmetries  may  allow to check the large
 $\im G_8$ results found at NLO, see for instance \cite{BP00,HPR03}
and references therein.
  
Moreover, it seems that some models beyond the Standard Model can achieve
values not much larger than $1\times 10^{-4}$, see for 
instance \cite{DIM00}.
It  will be important to check if  that value can be attained
in the Standard Model too. 

We are also finishing the analysis of the CP--violating asymmetries
in the integrated rates of the charged Kaon decays into three pions,
$\Delta \Gamma$ at NLO order in CHPT.

\section*{Acknowledgments}
This work has been supported in part by the RTN European Union
No HPRN-CT-2002-00311 (EURIDICE).
 E.G. is indebted to MECD (Spain) for a F.P.U. fellowship.
The work of E.G. and J.P. has been supported in part by MCYT (Spain)
 under Grants No. FPA2000-1558 and HF2001-0116 and  
by Junta de Andaluc\'{\i}a Grant No. FQM-101. The work of I.S. has been supported in part 
by the Swiss National Science Foundation and RTN, BBW-Contract 
No. 01.0357.

\end{document}